\documentclass{aa}
\usepackage{graphicx}
\usepackage{natbib}
\usepackage{txfonts}

\usepackage{hyperref}

\newcommand{\bc}{\begin{center}} 
\newcommand{\ec}{\end{center}}

\begin{document}

\title{Magnetic  flux cancellation in a flux-emergence magnetohydrodynamics simulation of coronal hole eruptions and jets}

\author{Spiros Patsourakos 
\and Vasilis Archontis  
}
\institute{Physics Department, University of Ioannina, Ioannina GR-45110,
Greece\\
\email{spatsour@uoi.gr}
}
\authorrunning{}
\titlerunning{Flux cancellation in 3D MHD emergence}

\date{Received  ; accepted }

 \abstract
{Observations have demonstrated that magnetic flux cancellation can be 
associated with coronal jets and eruptions taking place in coronal holes. However, magnetic flux cancellation is barely 
reported in magnetohydrodynamics (MHD) simulations of coronal jets and eruptions which employ emerging  twisted flux tubes.}
{We search for signatures of 
magnetic flux cancellation in a 3D resistive MHD flux-emergence simulation of coronal jets  and eruptions in a coronal-hole-like environment.}
{To do this, we analysed the output from a 3D MHD simulation of an emerging twisted horizontal flux tube
from the convection zone into the solar atmosphere. The simulation
considered the impact of neutral hydrogen on the magnetic induction equation, 
that is,  it 
employed partially ionised plasma. Standard
and blowout jets as well as  eruptions were observed during the simulation.}
{We observe clear evidence of magnetic flux cancellation in a short segment along
the internal polarity-inversion line (iPIL) of the photospheric $\mathrm{B_z}$ during an extended  period of the simulation characterised by eruptions 
and blowout jets. Converging magnetic footpoint motions at $\approx$ 1 
$\mathrm{km \, s^{-1} } $  carried sheared fields within the magnetic tails of the emerging flux tube towards the iPIL. These fields reconnect at the iPIL and generate concave-upward and slowly rising   field lines causing  a flux decrease that is associated with magnetic flux cancellation. 
The magnetic flux decreases
at a rate of $\approx 3.2 \times {10}^{18} $ $  \mathrm{Mx \, {hour}^{-1}}  $ and about 15-20 $\%$ during
intervals of individual eruptions and jets.}
{We show evidence of magnetic flux cancellation in 3D MHD simulations
of coronal hole eruptions and  jets associated with an emerging  twisted flux tube. The magnetic flux cancellation can be traced
up to about 520 km above the photosphere and might contribute to the formation of pre-eruptive magnetic flux rope seeds. 
Although our results are consistent with several basic aspects of 
magnetic flux-cancellation observations associated with coronal jets, the observations nevertheless 
also suggest that cancellation involves much larger fractions of the available flux than our numerical simulation. We supply avenues
to address this discrepancy in future work.}
\maketitle
\section{Introduction}
Coronal jets are narrow transient collimated outflows that are
observed all over the Sun: in the quiet Sun (QS), in coronal holes (polar and
equatorial), and in active regions \citep[e.g. see the reviews][]{2016SSRv..201....1R,2021RSPSA.47700217S,2022FrASS...920183S}.
They are mainly observed in the soft X-rays (SXRs) and in the extreme 
ultraviolet (EUV) with imagers and spectrometers
and in the white light with coronagraphs. It is important to build a concrete 
understanding of the physical mechanisms that form and
evolve coronal jets and small-scale eruptions in general because these phenomena might represent
an important contributor to the solar wind mass supply \citep[e.g.][]{2023ApJ...945...28R,2023Sci...381..867C}. In addition, given the relative
simplicity of the magnetic configurations involved in coronal jets, studying them and eventually
understanding them may also provide important clues about large-scale eruptive
solar phenomena \citep[e.g.][]{2015Natur.523..437S,2017Natur.544..452W,2021ApJ...907...41K}. 
Significant changes in the magnetic helicity and free magnetic 
energy,  similarly to what is observed during larger-scale eruptive phenomena, are recorded
during coronal jets, as suggested by observations and 
Magnetohydrodynamics (MHD) modelling \citep[e.g.][]{2024A&A...689L..11N,2024A&A...690A.181M}.
The multitude of
wavelength bands and instruments used  in observations of coronal jets
and related phenomena gives rise 
to a plethora of various descriptive terms and classifications.

A proposed classification of coronal jets is in terms of "standard" and "blowout"
jets \citet{2010ApJ...720..757M}. A standard
jet occurs when reconnection between an emerging magnetic bipole and the ambient  large-scale 
open magnetic field takes place \citep[e.g.][]{1996PASJ...48..353Y, 2008ApJ...673L.211M, 2013ApJ...769L..21A,2013ApJ...771...20M,2014ApJ...789L..19F}. 
Blowout jets occur when stressed and occasionally twisted
core magnetic fields first erupt and then reconnect with the ambient large-scale open magnetic field
\citep[e.g.][]{2009ApJ...691...61P,2013ApJ...769L..21A}. 
Therefore, a key characteristic of blowout jets is that 
a small-scale eruption is in order, and leads to the  opening of 
previously closed sheared or twisted core fields that are channelled into the resulting blowout jets.

Magnetic flux cancellation is a frequently observed phenomenon in the photosphere of
the QS, coronal holes, and active regions. It occurs
when patches of vertical photospheric magnetic fields of opposite polarity with a similar magnetic field magnitude approach each other
and their magnetic flux decreases \citep[][]{1985AuJPh..38..855L,1985AuJPh..38..929M,1988SoPh..117..243M}.
Frequently, periods of magnetic flux cancellation occur in tandem 
with multi-scale eruptive activity that ranges all way from proper coronal mass ejections
(CMEs) down to coronal jets and transient outflows 
\citep[e.g.][]{2009ApJ...700L..83G, 2012ApJ...759..105S, 2018ApJ...866....8Y, 2018ApJ...853..189P, 2019ApJ...871...67C,2022ApJ...933...12M}. In addition to being considered a potential trigger mechanism for eruptive phenomena 
and transient outflows and a  supplier of solar wind mass and 
coronal heating \citep[e.g.][]{2024ApJ...960...51P},
magnetic flux cancellation is also  a viable formation mechanism
of pre-eruptive magnetic flux ropes (MFRs; 
e.g. \citet{1989ApJ...343..971V,2003ApJ...595.1231A,2006ApJ...641..577M,2010ApJ...708..314A,2022ApJ...929L..23H,2024ApJ...966...70X}).
MFRs are coherent arrangements of  twisted magnetic fields that coil around
a common axis \citep[e.g. see the definitions and pertinent discussions in][]{2020SSRv..216..131P}, and they are
a candidate pre-eruptive magnetic configuration. 

There exist three possibilities to deal with the decrease in the magnetic
flux that is associated with magnetic flux cancellation:
i) emergence of U-loops; ii) submergence of $\Omega$-loops, and
iii) magnetic reconnection at or close to the photosphere that creates pairs of U- and $\Omega$-loops
(for a schematic of these three cases, we refer to Fig. 2 in \citet{1987ARA&A..25...83Z}). The third scenario generates both   U- and $\Omega$-loops that emerge and submerge, respectively.
It then depends on whether the height at which the reconnection occurs is somewhat below or above the height at which
magnetic field information is available, so that magnetic flux cancellation is detected via emerging
U- or  submerging $\Omega$-loops, respectively.
Observations of magnetic flux cancellation give rise to 
cases consistent with either emerging U-loops
\citep[e.g.][]{2000A&A...364..845V, 2002SoPh..209..119B,2005ApJ...626L.125B} 
or submerging $\Omega$-loops \citep[e.g.][]{1999SoPh..190...35H,2004ApJ...602L..65C,2009ApJ...703.1012Y, 2012SoPh..281..599T}.

Observations by the Atmospheric Imaging Assembly (AIA) and Helioseismic and Magnetic Imager (HMI) on board
the Solar Dynamics Observatory
(SDO)  showed varying proportions of coronal jets and small-scale eruptions associated
with magnetic flux cancellation. \citet{2017AGUFMSH43A2796M} analysed 
30 equatorial coronal hole (ECH) and 30 QS jets and found that 85 \% of them
were associated with magnetic flux cancellation.
\citet{2018A&A...619A..55M} studied
21 small-scale eruptions stemming from 11 QS coronal bright points (CBPs) and found that 19 of these eruptions were associated
with magnetic flux cancellation. \citet{2018ApJ...853..189P} showed 
that magnetic flux cancellation was associated with 13 ECH jets containing mini-filament material. \citet{2022ApJ...933...12M} analysed
43 microflares in ECHs that were associated with small-scale bipolar ephemeral
active regions with a typical span of 10 Mm. 
18 of the microflares were associated with blowout jets,  and magnetic flux cancellation before and during those jets was reported. Recent Solar Orbiter Extreme-Ultraviolet Imager (EUI) ultra-high spatial resolution
observations of five small-scale coronal jets showed that three of them
were 
associated with
magnetic flux cancellation \citep{2023ApJ...943...24P}.
These studies also suggest that magnetic  flux cancellation might be
the trigger of the reported eruptions and jets. 
On the other hand, \citet{2019ApJ...873...93K} analysed 27 ECH 
jets and found that only 6 of them 
were associated with magnetic flux cancellation that occurred before
or during them. \citet{2021ApJ...909..133M} 
applied local correlation tracking to white-light photospheric images at the 
feet of a sample of 35 ECH jets and 
found evidence of converging motions and magnetic flux cancellation
in four cases, and eight more complex cases, featuring both
flux emergence and magnetic flux cancellation. Finally, recent
joint EUI-HMI observations of  627 small-scale QS brightenings 
that occurred above
concentrations of strong bipolar magnetic fields showed that only about
8 \%  of these events were associated with magnetic flux cancellation \citep{2024A&A...692A.236N}.
The smallest magnetic flux   emergence or cancellation events might be missed by the HMI resolution, as was noted for example by \citet{2024A&A...686A.218N,2025geor}.
A recent spectroscopic study of two jets in an ECH showed no evidence of magnetic flux cancellation in HMI observations \citep{2024A&A...690A..11K}.  

The MHD modelling of coronal jets and of eruptive solar
phenomena  in the low solar atmosphere in general may be cast into two broad categories: surface-driven and  flux-emergence models \citep[e.g. consult the recent reviews of][]{2011LRSP....8....1C,2016SSRv..201....1R,2018SSRv..214...46G,2019RSPTA.37780387A,2020SSRv..216..131P,2021RSPSA.47700217S, 2022FrASS...920183S}.
Surface-driven models employ simulation boxes
that only incorporate the solar atmosphere, and they therefore span 
the photosphere, chromosphere, the transition region, and a segment
of the lower corona.  These models are driven
by prescribed
photospheric flow fields.
Flux-emergence models typically include a part of the upper
convection zone along with
a segment covering the stratified atmosphere above it. These models are driven by
the emergence of buoyant  horizontal or toroidal twisted flux tubes from
the convection zone into the solar atmosphere. In addition to prescribing
the specifics (e.g. distribution of twist and a
density deficit) of the sub-photospheric flux tubes, the subsequent evolution
of the system including emergence, build-up of magnetic energy, shear and twist and 
eruptive activity is rather self-consistent. 

Magnetic flux cancellation in surface-driven models and simulations
was addressed by various authors \citep[e.g.][]{1989ApJ...343..971V, 2003ApJ...595.1231A, 2006ApJ...641..577M, 2010ApJ...708..314A,2019ApJ...872...32S,2023A&A...675A..97F,2024ApJ...960...51P}. 
In order to achieve magnetic flux cancellation, these models and simulations typically involved
a combination of postulated converging and  shearing
photospheric flow fields along with magnetic
diffusion of the normal component of the magnetic field via prescribed tangential electric fields. Imposing appropriate sub-photospheric 
velocity fields was  also shown to give rise to magnetic flux cancellation \citep{2023ApJ...955..105R}.

Magnetic flux cancellation in flux-emergence simulations
was discussed in detail by \citet{2011PASJ...63..417M} and \citet{2012ApJ...754...15F}.
Both studies involved the emergence of a horizontal
twisted flux tube from the convection zone into the solar atmosphere.
\citet{2011PASJ...63..417M} reported that
magnetic flux cancellation was caused by the 
emergence and subsequent convergence of U-loops in the related polarity-inversion
line (PIL) of the photospheric $B_{z}$,  where
$B_{z}$ changes sign when the PIL is crossed. The U-loops might emerge against the gravity
of the plasma collected in their concave upward segments, that is,  magnetic 
dips, if they are shallow enough so that mass can drain. This essentially facilitates the rise of  
the U-loop. However, deep U-loops are not able not
rise, and their legs would therefore be pinched off and reconnect.  
In both cases, signatures of magnetic flux cancellation are expected. 

\citet{2012ApJ...754...15F} reported that
photospheric convective and shearing  motions carried 
emerging  magnetic fields of opposite polarity in the PIL where
they reconnected in a tether-cutting fashion \citep[e.g.][]{2001ApJ...552..833M}.
The reconnection gave rise to magnetic flux cancellation
signatures, and an MFR formed. None of these simulations reported eruptive activity. 
To the best to our knowledge, the studies by \citet{2011PASJ...63..417M} and \citet{2012ApJ...754...15F} are the only 3D MHD simulations
of emerging twisted flux tubes that reported a magnetic flux cancellation.
\begin{figure}[!h]
\centering
\includegraphics[width=0.5\textwidth]{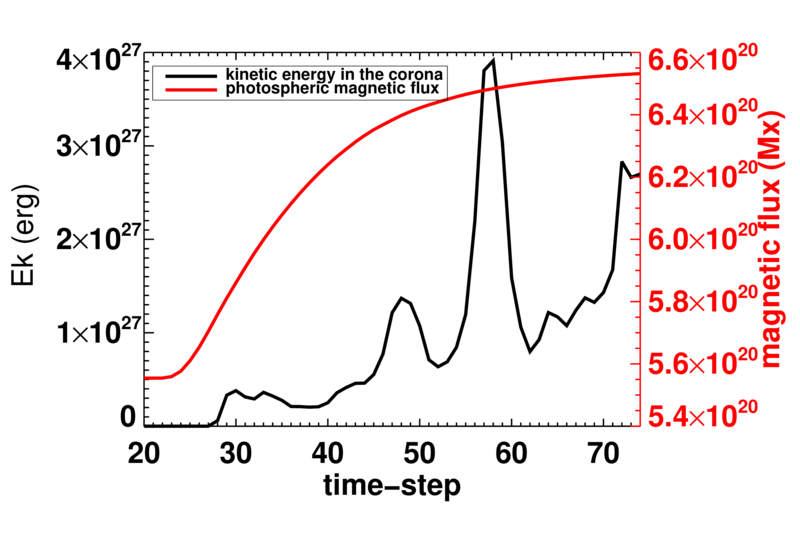}
\caption{Temporal evolution of the coronal kinetic energy calculated 3.2 Mm above the photosphere (black line) and of the photospheric magnetic flux (red line).  Consecutive time-steps are 86.9 s apart.}
\label{fig:ke}
\end{figure}

\citet{2022ApJ...935L..21N} discussed 2D Bifrost flux-emergence simulations
in a magnetic-null configuration within a coronal hole and studied the formation and dynamics
of a CBP in terms of microflares, eruptions, and jets.  In this simulation, flux emergence is related
to convection-mediated dynamics. 
The authors noted that the photospheric magnetic flux at the feet of the CBP
decreased for a time, which they attributed to magnetic flux cancellation.  
\citet{2023ApJ...943...24P} reported Bifrost simulations
of the injection of a horizontal flux sheet with a temporally varying
strength from the convection zone into the solar atmosphere. They analysed five small-scale jets in detail and found that four of them were triggered by magnetic flux cancellation. The reported jets contained cool plasma. The simulations by \citet{2022ApJ...935L..21N} and \citet{2023ApJ...943...24P} did not employ emerging twisted flux tubes.

We therefore conclude that there is a  lack of detailed investigations of magnetic flux cancellation in MHD simulations of emerging twisted flux tubes that lead to coronal jets and eruptions in coronal holes. 
This is the focus of our study and  a  timely investigation, also in view of  jet observations showing evidence of magnetic flux cancellation. In Section 2 we briefly describe our simulation, Section 3 presents the overall behaviour of our simulation,
Section 4 discusses the characteristics and 
origins of the detected magnetic flux cancellation in detail and describes its implications for pre-eruptive MFRs, and
Section 5 finally contains a summary and a discussion of our results. It also supplies comparisons of our simulation results with pertinent observations and MHD models.
\begin{figure*}[!ht]
\centering
\includegraphics[width=0.99\textwidth]{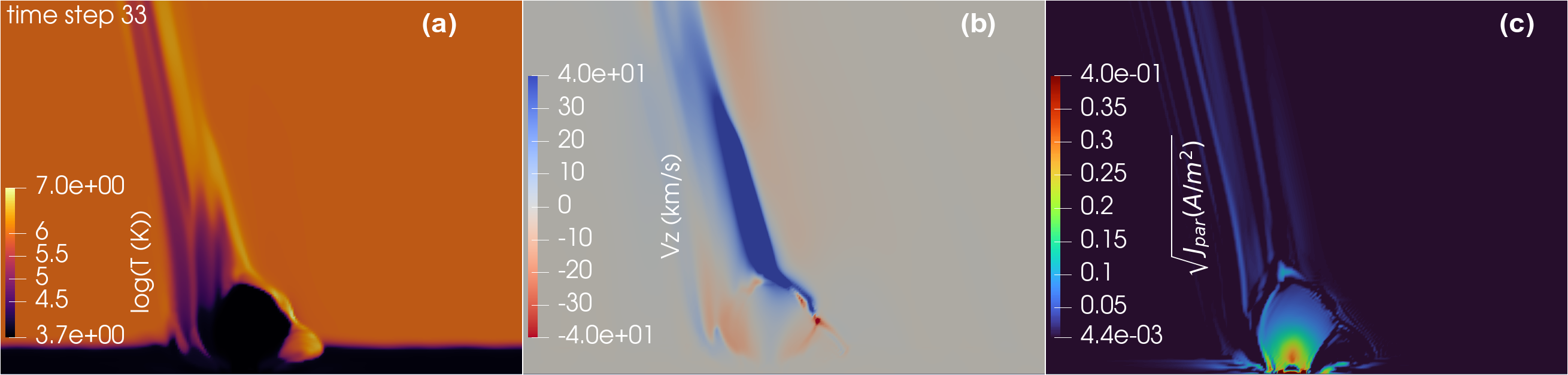} 

\includegraphics[width=0.99\textwidth]{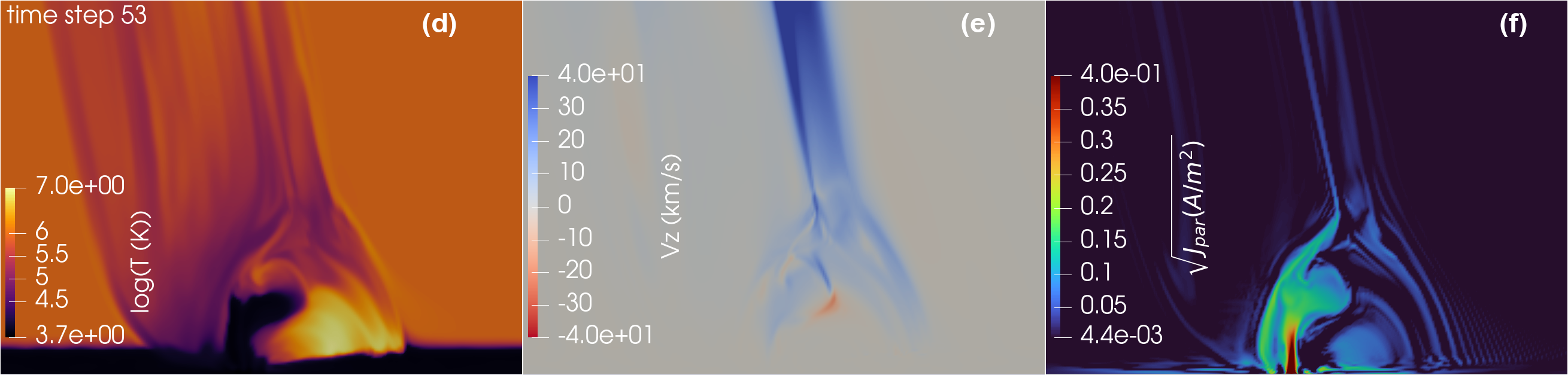} 

\includegraphics[width=0.99\textwidth]{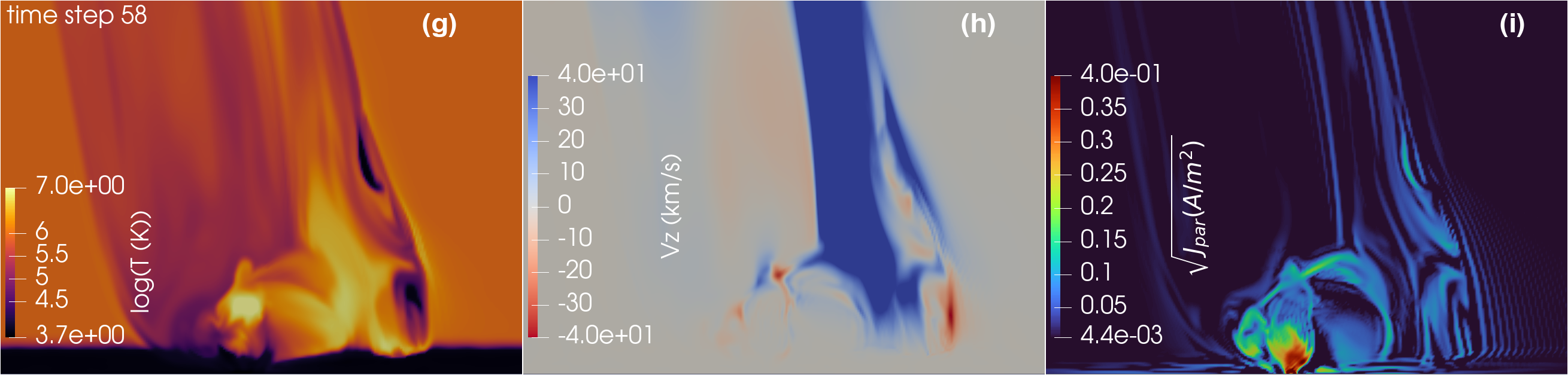}
\caption{Distribution of the logarithm of the temperature, $V_z$ and $J_{par}$ (square-root scaling) in the 
$xz$-midplane spanning the solar atmosphere (i.e. photosphere and above) in the left, middle, and right column, respectively. Each row corresponds to
a given snapshot during the simulation. Consecutive time-steps are 86.9 s apart. The associated movie is available online.}
\label{fig:jpar}
\end{figure*}

\begin{figure}[!h]
\centering
\includegraphics[width=0.5\textwidth]{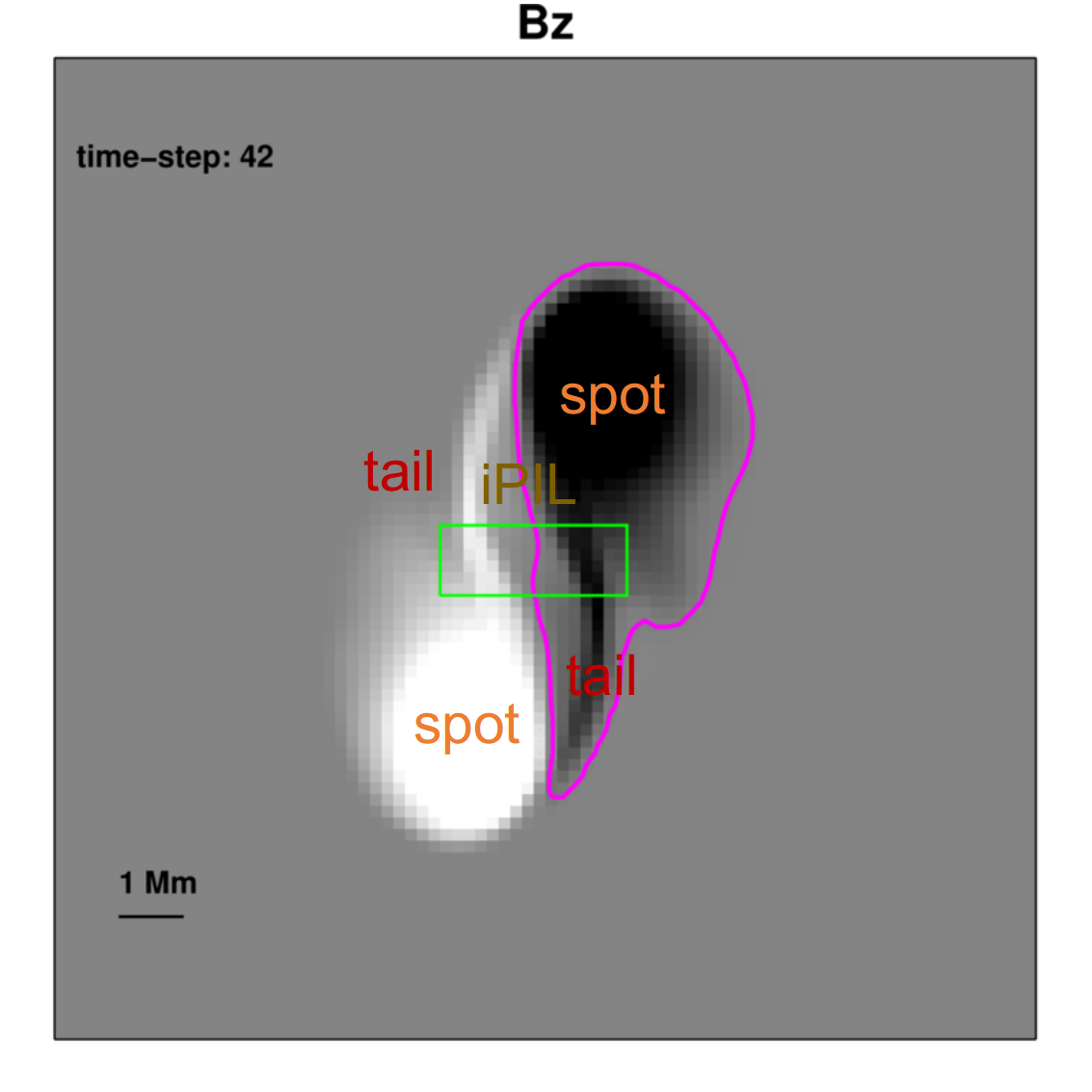}
\caption{Photospheric $B_{z}$ for time-step 42. Black (white) corresponds to
negative (positive)  $B_{z}$  saturated
in $\pm$ 300 Gauss. The purple line corresponds
to the PIL of the photospheric $B_z$. The green box
encapsulates the region in which magnetic flux cancellation
takes place.}
\label{fig:bz42}
\end{figure}

\begin{figure*}[!h]
\centering
\includegraphics[width=0.9\textwidth]{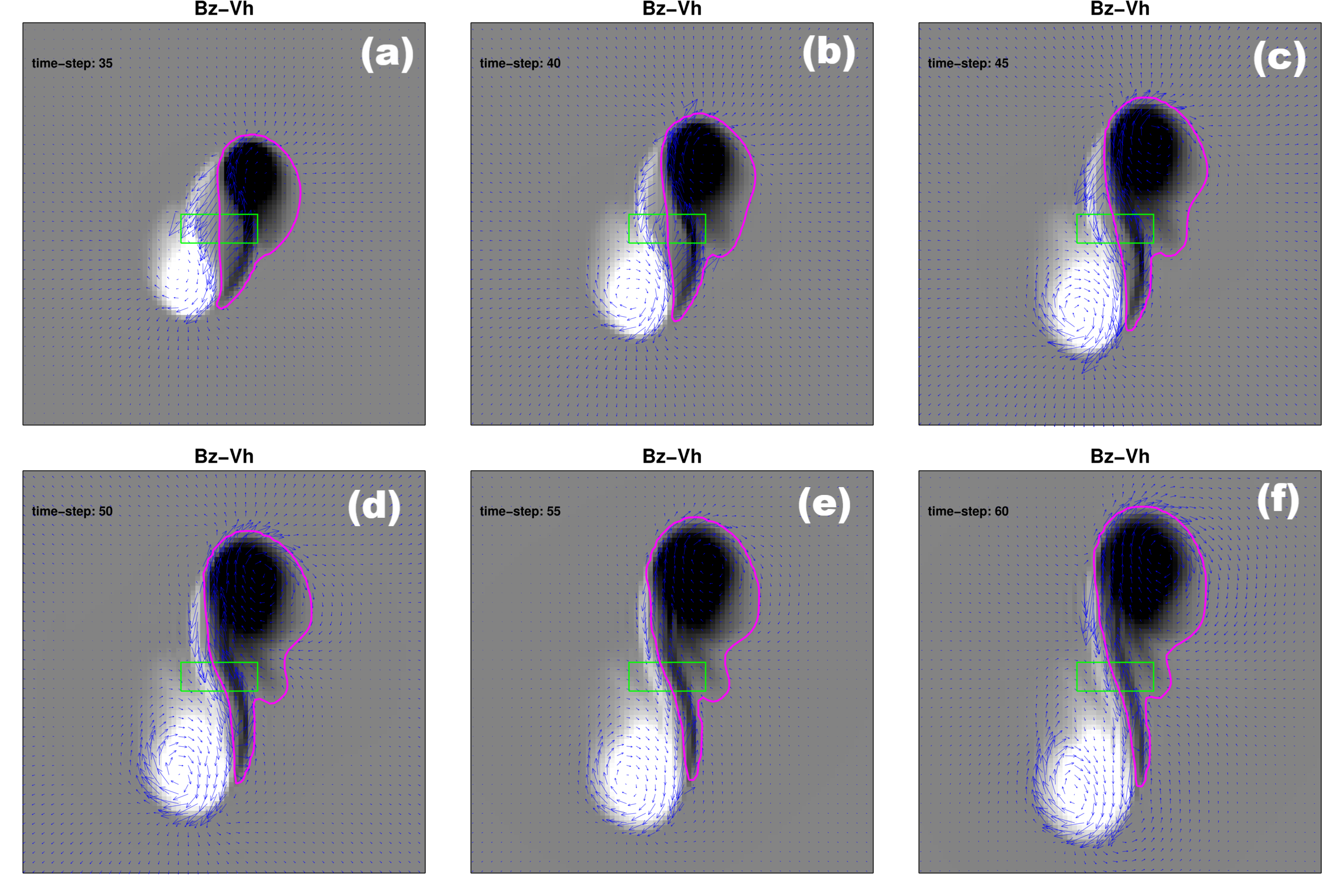}
\caption{Sample panels of the 
photospheric $B_{z}$ and of the magnetic-footpoint horizontal  speeds. Black (white) correspond to
negative(positive) polarity $ B_{z} $ saturated
in $\pm$ 300 Gauss. The overplotted blue arrows
correspond to the horizontal speeds of the magnetic footpoints. The longest arrows
correspond to a magnitude of 1 $\mathrm{km \, s^{-1} } $.
The purple line corresponds
to the PIL of the photospheric $ B_{z} $. The green box
corresponds to the region in which magnetic flux cancellation
takes place.  Consecutive time-steps are 86.9 s apart. The associated movie is available online.}
\label{fig:montvh}
\end{figure*}

\begin{figure}[!h]
\centering
\includegraphics[width=0.5\textwidth]{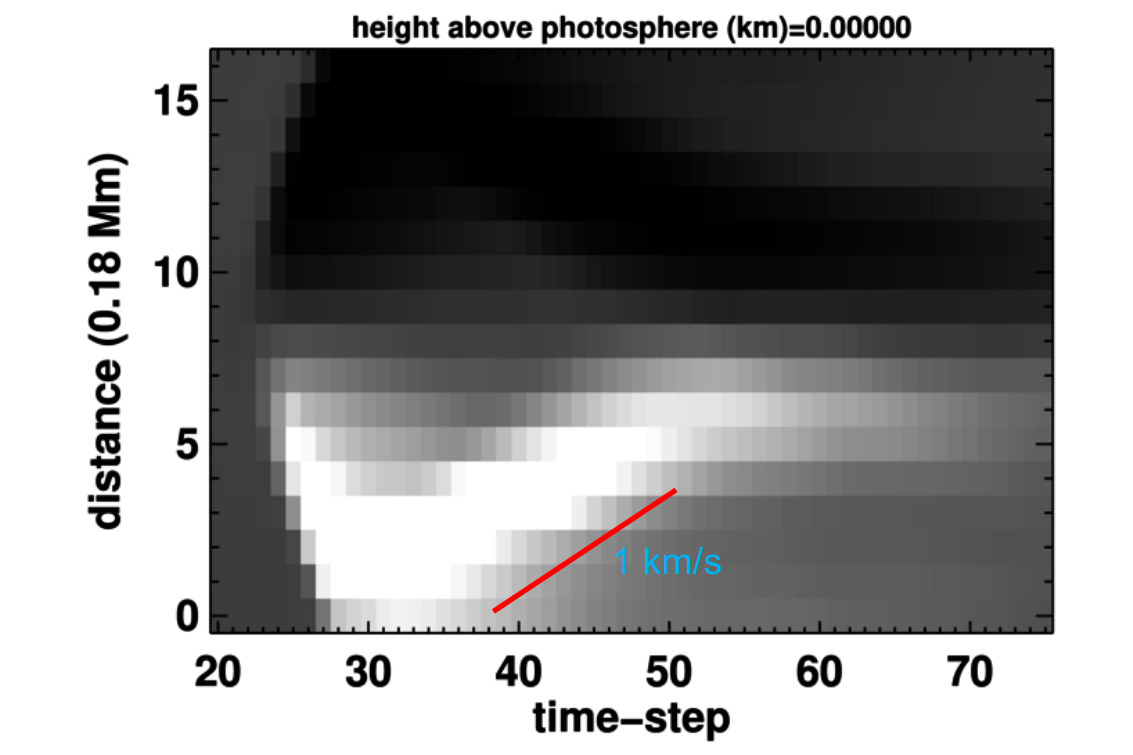}
\caption{Time-distance plot of the photospheric   $B_{z}$ corresponding to sums of $B_{z}$ along the columns of the green box
of Fig. \ref{fig:bz42}; the distance corresponds to
the position along the base of the box of Fig. \ref{fig:bz42}.  The inclined
red line has a slope of  $\approx$ 1 $\mathrm{km \, s}^{-1}$. Consecutive time-steps are 86.9 s apart.
}
\label{fig:stack}
\end{figure}

\section{Simulation}
We employed output from
a flux-emergence simulation
using the code Lare3D \citep{2001JCoPh.171..151A}, which solves the single-fluid ressistive  MHD  equations in 3D.
The simulation, reported by \citet{2023ApJ...952...21C}, employed partial ionisation of the hydrogen
in the magnetic induction equation. 
For more details on the implementation and the overall impact of partial ionisation
of hydrogen 
in flux emergence in 3D MHD simulations, we refer to \citet{2023ApJ...952...21C}.
The same simulation was also recently analysed by \citet{2024A&A...690A.181M} , who
studied the free magnetic energy and magnetic helicity evolution in the simulation box and its relation to coronal jets and eruptions.
The simulation box was a ${420}^{3}$ voxel cube. The sides of each
voxel were 180 km long, and therefore,
the simulation box spanned about 65 Mm in each dimension. The simulation box included an $\approx$ 7 Mm thick
upper convection zone layer at its bottom, and photospheric, chromospheric, transition region, and coronal layers
were also included. The photospheric layer corresponds to the $xy$ plane   of the simulation box at an appropriate height.
A cylindrical and uniformly twisted flux  tube was placed 2.3 Mm below the photosphere with its axis running along the $y$ -axis of the simulation box.
Given an ascribed density deficit along a segment
of the flux tube, the tube becomes buoyant and eventually emerges
into a background isothermal atmosphere. The background atmosphere  is  permeated with
an almost vertical magnetic field
that forms an angle of $\approx$ 11 degrees with the vertical and uniform positive-polarity ambient magnetic field having a magnitude of $\approx$ 10 Gauss.
In this way, our simulation emulated the emergence of a bipolar region within a coronal hole-like environment. The results from the simulation were damped in 86.9 s time-steps. We stress that the emergence of the twisted flux tube into the solar atmosphere is not full, but partial, that is,  the tube axis
does not   emerge above the photosphere (see \citet{2023ApJ...952...21C}). This behaviour is due to the mass-laden lower
segments of the buoyant  tube, where plasma is collected in magnetic dips. This inhibits the rise of the tube. The partial emergence of twisted flux tubes is a  standard feature of many flux-emergence simulations
\citep[e.g. consult the review][]{2019RSPTA.37780387A}.

\section{Overall behaviour of the simulation}

We first discuss the overall behaviour of the simulation.
In Fig.\ref{fig:ke} we plot the temporal evolution
of the total unsigned photospheric magnetic flux and the kinetic energy calculated in the coronal segment of the simulation box, that is,   starting 3.2 Mm above the photosphere. The photospheric magnetic flux increased during
the considered interval: rapidly from time-steps 25 to 50, and more gradually from time-step 50 onward. This behaviour is due
to the emergence of the twisted flux tube from the convection zone into the solar atmosphere, and it is typical of either simulations or observations
of a flux emergence in the solar atmosphere.
The kinetic energy exhibits several major temporally resolved peaks at
time-steps 30, 34, 48, 58, and 75; a bump in kinetic energy is also seen  at  time-step 45.
As discussed in detail by \citet{2024A&A...690A.181M},
kinetic energy peaks  until $\approx$ time-step 34 correspond
to standard (i.e.  reconnection) jets, while for the rest of the considered simulation interval, they correspond
to blowout jets caused by eruptions
of stressed  magnetic fields.  The standard jets of our simulation involve cool surge-like
plasmas, as also seen in other simulations \citet{1996PASJ...48..353Y, 2013ApJ...771...20M, 2016ApJ...822...18N}. More cool material is involved in the blowout jets of our simulation.

Fig.\ref{fig:jpar} contains slices of the temperature,
vertical speed ($V_z$), and field-aligned 
current density, $J_{par}$,
\begin{equation}
J_{par} =\frac{\mathbf{J} \cdot \mathbf{B}}{|\mathbf{B}|},
\end{equation}
in the $xz$-midplane spanning the solar atmosphere (i.e.  photosphere and above)  for three snapshots during the simulation. Because the axis of the emerging flux  tube runs along the $y$-axis, slices of physical parameters in the $xz$-midplane  sample a  cross-section perpendicular to and in the middle of the emerging flux tube axis.
In  the first row of  Fig.\ref{fig:jpar},  we show a snapshot during a
standard jet, while the second and third row of this figure  
contain a snapshot of a pre-eruptive MFR and its eventual eruption into a blowout jet, respectively.
The snapshot in the second row of Fig.\ref{fig:jpar} corresponds to
a period of almost constant coronal kinetic energy (see Fig.\ref{fig:ke}), before it impulsively
increases during an eruption associated with a blowout jet. We discuss this pre-eruptive MFR in more detail in Section 5.
The standard jet (first row) corresponds to an extended region with high $V_z$ values (panel (b)) that coincides with 
a high-temperature region (panel (a)). These are telltale signatures of standard jets that occur when reconnection
between   emerging closed and   ambient open fields takes place. When  these two sets
of field lines start to reconnect, the heated plasma is emitted in the form of the standard jet.
In the second row of Fig.\ref{fig:jpar}, 
the cross-section of a pre-eruptive MFR can  be traced by a quasi-circular concentration in  $J_{par}$ (panel (f)), and it mainly contains
cool plasma (panel (d)). The eruption of this MFR generates a large blowout jet (panel (h)) that contains both cool and hot plasma (panel (g)).
Inspection of the associated movie also shows a multitude of 
quasi-steady upflows in addition to the larger-scale eruptions and jets.  

In Fig.\ref{fig:bz42} we show the photospheric $B_{z}$ for time-step 42. In this snapshot, a significant part of the flux emergence has taken place 
(see the photospheric magnetic flux evolution in Fig.\ref{fig:ke}).
The PIL  of the photospheric $B_{z}$ (purple line) 
consists of two parts: one part between the two  polarities 
of the emerging magnetic flux (inner PIL; iPIL) and the other part 
lies around the minority (negative) polarity emerging
flux and essentially separates it from the ambient positive-polarity field. 
The photospheric  imprint of the emerging  magnetic flux tube
consists of two  components at either side of the iPIL:
two oval-shaped  strong-field  spots,
and two elongated weaker-field    magnetic tails 
corresponding 
to the feet and the horizontal extensions of the  emerging flux tube
into the photosphere, respectively (e.g. \citet{2000ApJ...544..540L,  2010A&A...514A..56A, 2011SoPh..270...45L}).
Along most of the length of the iPIL, it separates spot-tail segments with a disparate
magnetic field magnitude. However, around the centre of Fig.\ref{fig:bz42}, and encompassed by the green box, lies a  segment of the iPIL that features
similar
magnitude  conjugate tail magnetic fields. 
As we show in the next section, the magnetic flux cancellation
takes place in this region.

\begin{figure}[!h]
\centering
\includegraphics[width=0.5\textwidth]{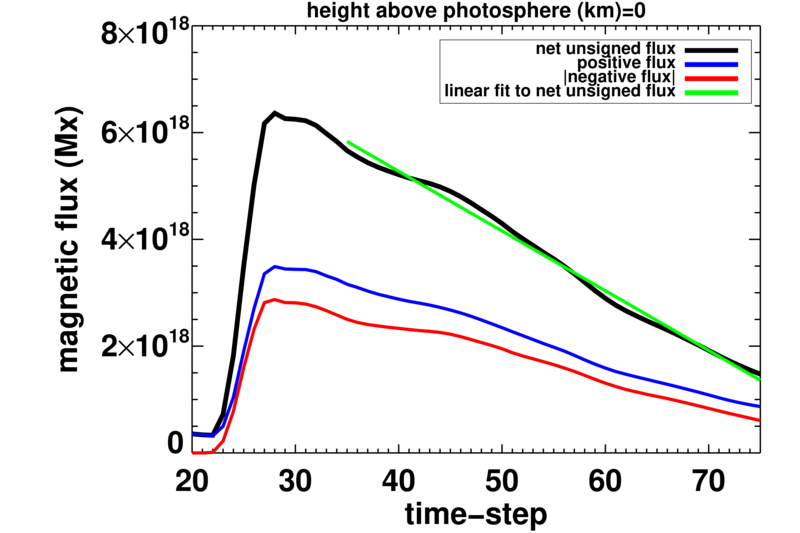}
\caption{Temporal evolution of the positive (blue line), negative (red line)
and net unsigned photospheric  magnetic flux (black line) in the green box of Fig.\ref{fig:bz42} shown from the appropriate column sums of the time-distance map of Fig.\ref{fig:stack}. The green line corresponds to a linear fit of the
time-step and net unsigned flux pairs  from time-step 35 onward. Consecutive time-steps are 86.9 s apart.}
\label{fig:jmapflux1}
\end{figure}

\begin{figure}[!h]
\centering
\includegraphics[width=0.5\textwidth]{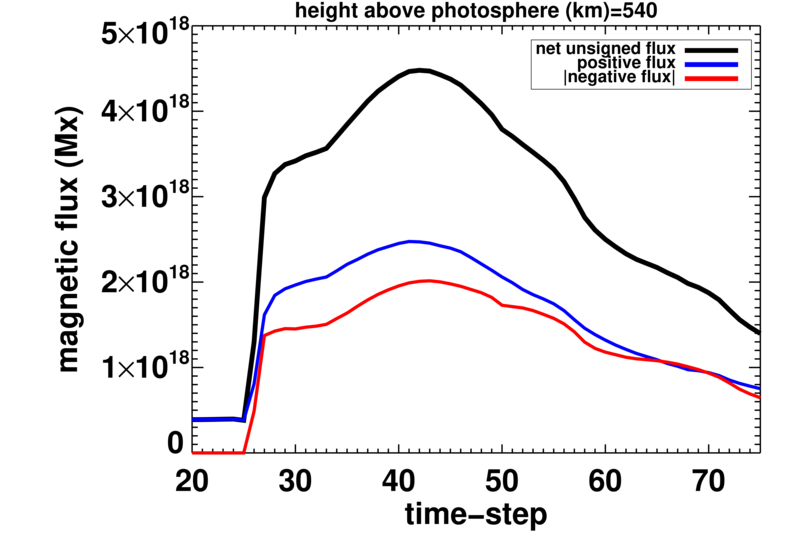}
\caption{Same as in Fig. \ref{fig:jmapflux1}, but this time, corresponding
to magnetic fluxes calculated 520 km above the photosphere and in the same
box as used for the calculations of Fig. \ref{fig:jmapflux1}.}
\label{fig:jmapflux2}
\end{figure}

\begin{figure}[!ht]
\centering
\includegraphics[width=0.5\textwidth]{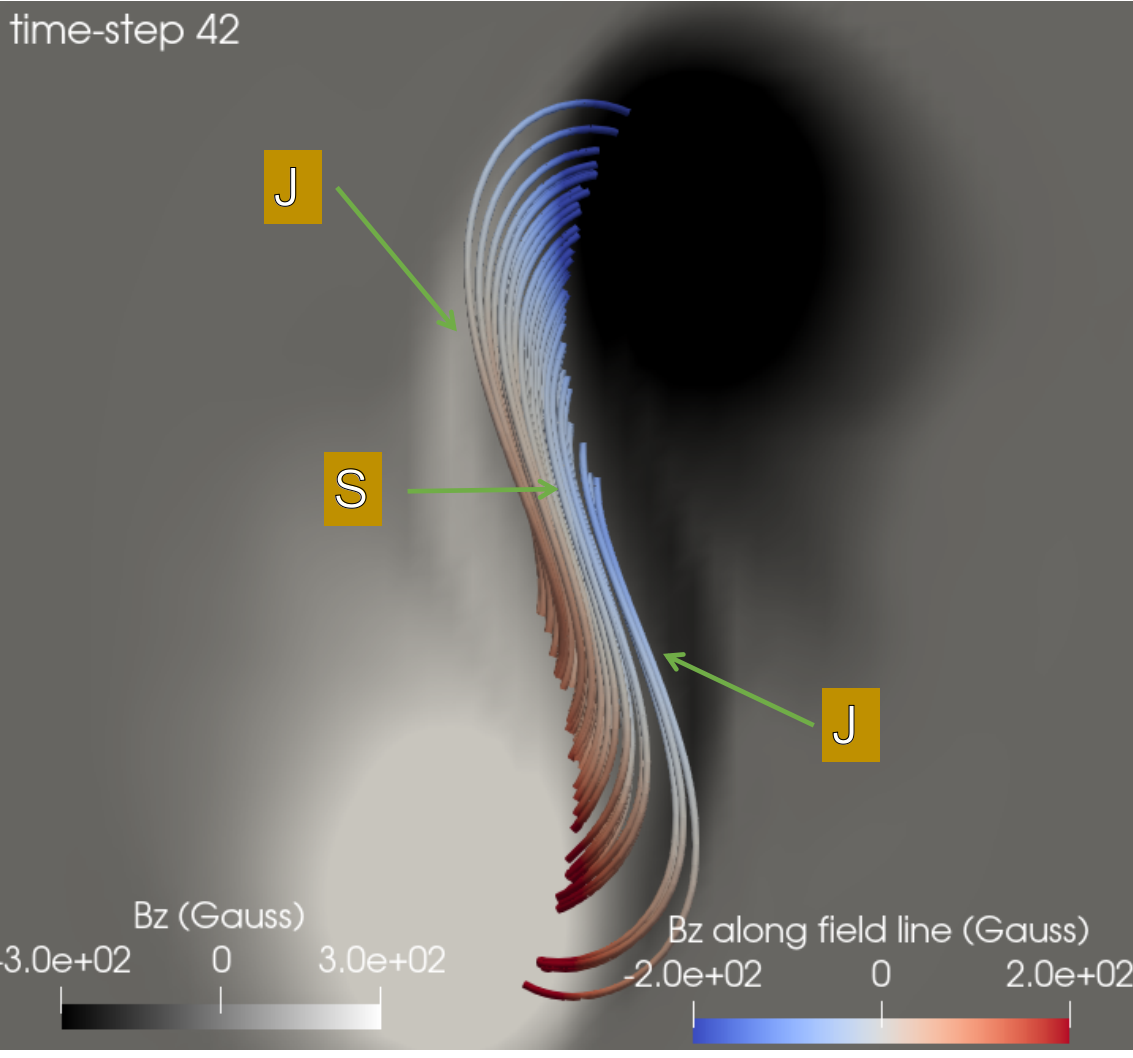}
\caption{Magnetic field lines traced from
a sphere with a radius of 2 pixels and the centre 2 pixels above the middle of the photospheric plane of the simulation box for time-step 42. We display the photospheric $B_{z}$ saturated in 
$\pm$ 300 Gauss with a greyscale. The traced field lines are coloured as a function of $B_{z}$ along their lengths.}
\label{fig:bzpilines}
\end{figure}

\begin{figure*}[!h]
\centering
\includegraphics[width=0.9\textwidth]{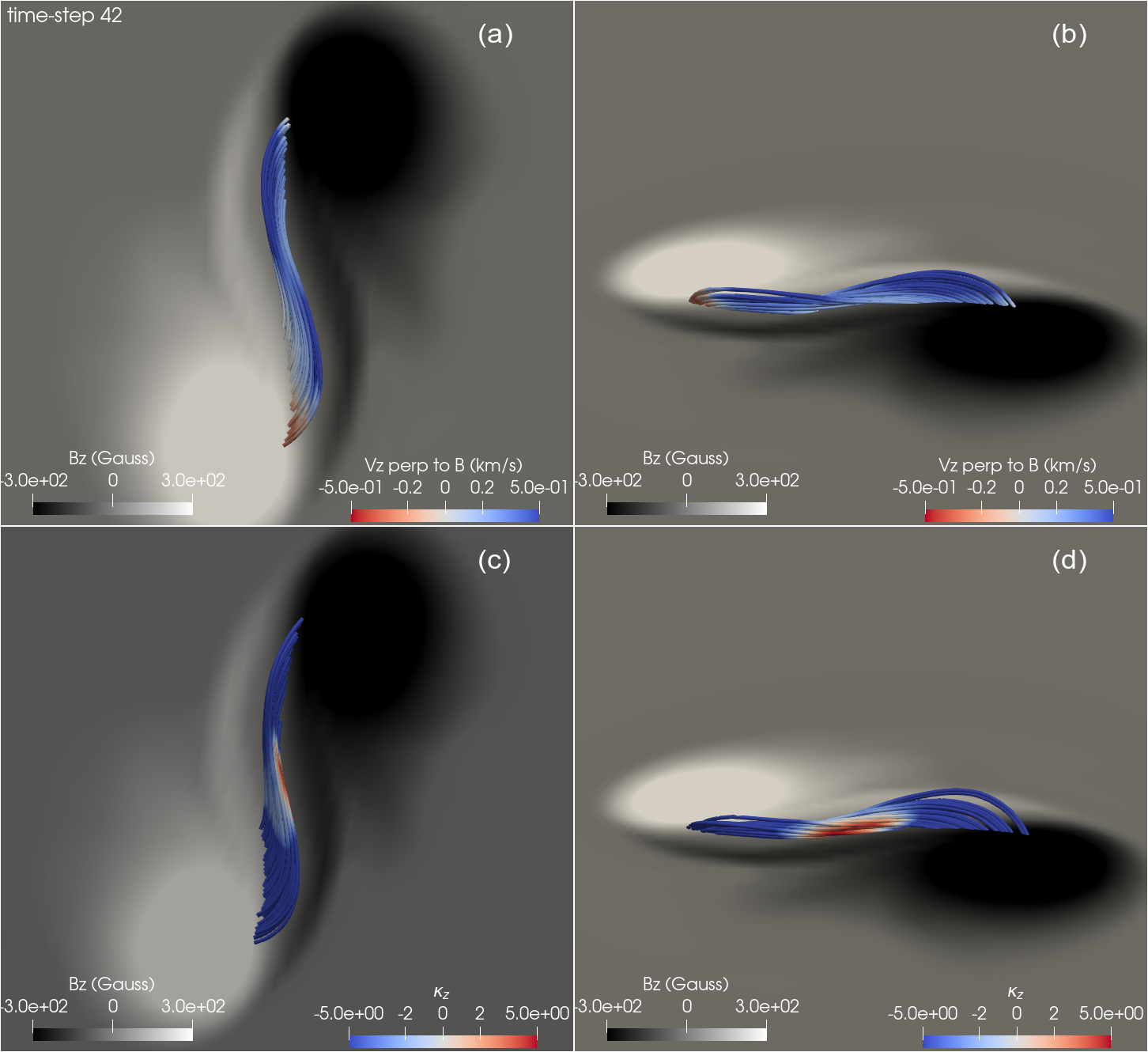}
\caption{S-shaped magnetic field lines of Fig. \ref{fig:bzpilines}  
coloured as a function of  $V_{z}^{\perp}$
(upper row) and $\kappa_{z}$ (lower row)  along their lengths for time-step 42. The left (right) column 
corresponds to overhead (side) views.}
\label{fig:pil}
\end{figure*}

\begin{figure}[!h]
\centering
\includegraphics[width=0.5\textwidth]{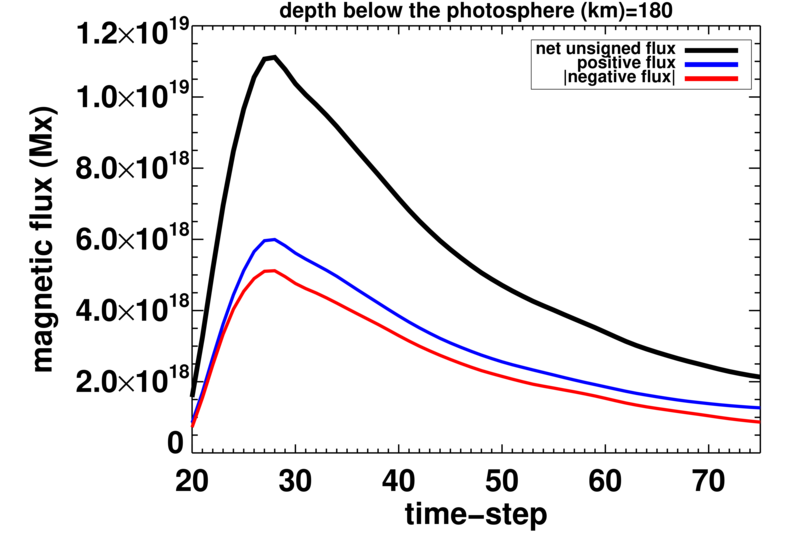}
\caption{Temporal evolution of the positive (blue line) and negative (red line) magnetic flux 180 km below the
photosphere for the green box of Fig.3. Consecutive time-steps are 86.9 s apart.}
\label{fig:jmapflux3}
\end{figure}

\begin{figure*}[!ht]
\centering
\includegraphics[width=0.9\textwidth]{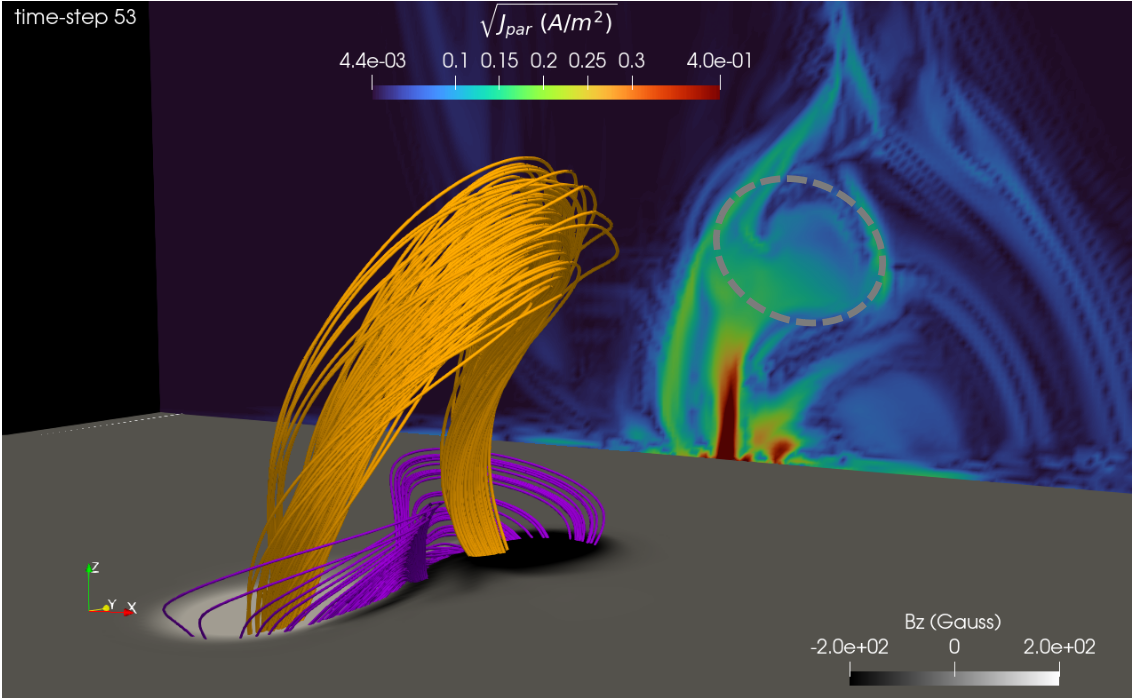}
\caption{Pre-eruptive MFR (yellow magnetic field lines)
traced 
from a small sphere centred at and spanning
a quasi-circular concentration of enhanced  $J_{par}$ outlined with
dashes in the $xz$-midplane. 
The $J_{par}$ distribution in the $xz$-midplane  is translated from its original position for clarity.
The purple magnetic field lines 
are traced from the same seeds as the field lines shown in Fig. \ref{fig:bzpilines}.
We display
the photospheric $B_{z}$ distribution with a greyscale. 
The figure corresponds to time-step 53.}
\label{fig:bzjparlines}
\end{figure*}

\section{Magnetic flux cancellation}
\subsection{Detection of magnetic flux cancellation}
In this section, we discuss the signatures and characteristics of magnetic flux cancellation
as they result from our simulation.
Fig.\ref{fig:montvh}  contains several representative panels of the photospheric
$B_{z}$ during the simulation, which can essentially be treated
as photospheric $B_{z}$ magnetograms. Each
panel in this figure also includes the horizontal velocities of the photospheric magnetic footpoints, which are shown with blue arrows. 
In determining these velocities, we followed
\citet{2003SoPh..215..203D} and
\citet{2020ApJ...891...14L}, who considered the impact
of the flux emergence on the horizontal velocities of the magnetic footpoints
(for an illustration of the effect, see, e.g., Figure 1 in \citet{2003SoPh..215..203D}).
This is a necessary action to take, as we hereby 
use a flux-emergence simulation, and 
residing exclusively to the horizontal (plasma) velocities
is not sufficient  for the calculation of the magnetic-footpoints' velocities. 

Fig.\ref{fig:montvh} and the associated movie reveal several  generic features of MHD simulations of emerging twisted flux tubes that lead to the formation of  bipolar regions with strong PILs. This includes 
the appearance and development of  
conjugate  magnetic spots
and tails
as well as complex flow fields with shear and rotation.
The spots
start to emerge at about time-step 23, and the tails appear slightly
later, at about time-step 26.
The $B_{z}$ evolution in the green box
(same as in
Fig.\ref{fig:bz42}) shows that
during time-steps $\approx$ 20-35, the magnetic
flux moves away from the iPIL due
to the emerging and laterally expanding
fields. However, at time-step $\approx$ 35,  the segments of the conjugate 
tails inside the green box  start  to converge towards the  iPIL  and  eventually reach a minimum 
distance at  about time-step 53. 
The convergence of the magnetic tails
is accompanied by a decrease in their magnetic field magnitude. The green
box in Fig.\ref{fig:bz42} and Fig.\ref{fig:montvh}  sufficiently covers the full length along the iPIL where the
tail-tail convergence and interaction described above take place. 

The magnetic footpoint velocities of Fig.\ref{fig:montvh} show an opposite
rotation of the two spots, which drags the conjugate magnetic tails  in opposite directions and hence  shears them.
These rotating and shearing motions
result from Lorentz forces, as in many other MHD simulations and models
\citep[e.g.][]{2000ApJ...545.1089L,2001ApJ...547..503M,2008A&A...492L..35A,2009ApJ...697.1529F,2012ApJ...754...15F, 2012ApJ...761...61G,2013ApJ...778...99L,2015A&A...582A..76S,2017ApJ...850...95S}.
As the iPIL is inclined with respect to the $x$-axis, and because of
the  opposite-directed shearing footpoint motions on either side of the iPIL, we  surmise that these motions introduce a component perpendicular to
and pointing to the iPIL. This essentially causes 
the conjugate magnetic tails
to converge  to  the iPIL.

To furthermore study the evolution of the conjugate magnetic tails, we generated
Fig.\ref{fig:stack}, where we display
a time-distance plot of  $B_{z}$ averaged along the height of 
the green box in Fig.\ref{fig:bz42} 
as a function of time and distance along the base of the box.
During time-steps $\approx$ 23-35, we observe two diverging branches 
of opposite-signed  $B_{z}$.  Starting from time-step 35, however, the
two branches start to converge and then attain a more or less
constant distance at about time-step 53. The speed at which the two
branches converge is around 1 $ \mathrm{km \, s^{-1} } $e judged by the fiducial red line
we supply in Fig.\ref{fig:stack}. These patterns  are consistent
with the photospheric magnetograms
we discussed in the previous paragraphs.

In Fig.\ref{fig:jmapflux1} we show time-series of the positive, absolute  negative, and net unsigned magnetic
flux calculated for the green box in Fig.\ref{fig:bz42}. This is equivalent to taking
column sums of the time-distance plot in Fig.\ref{fig:stack}.
We first note that the positive and negative fluxes are relatively well balanced. The positive flux is slightly higher as a natural consequence of the positive-polarity ambient magnetic field. From time-step 35 onward, which corresponds to the period of the conjugate-tail convergence and encounter discussed
in the previous paragraphs, all fluxes decrease simultaneously. There is a significant
decrease of $\approx$ 75 \% in the net unsigned magnetic flux
during this period. This corresponds to a flux-cancellation rate of $\approx$  3.2 $\times {10}^{18}$ $  \mathrm{Mx \, {hour}^{-1}}  $ 
as resulting from the linear fit of the pertinent time-step and magnetic flux pairs (green
line in Fig.\ref{fig:jmapflux1}).
In  shorter intervals that encompass major peaks in the coronal kinetic energy that correspond to eruptions and  blowout
jets, for example, from about time-steps 45-53 and 55-62 in 
Fig. \ref{fig:ke}, the magnetic flux 
decreases by about 15-20 $\%$  of its value at the start of the corresponding
intervals.
In summary, Figures \ref{fig:bz42}-\ref{fig:jmapflux1}, and given
the pertinent results presented in the Introduction, show telltale
signatures of magnetic flux cancellation. This includes 
convergence followed by a decrease in their flux content in a PIL of 
photospheric magnetic patches of opposite polarity and 
similar magnetic field magnitude and flux.

Magnetic flux cancellation associated with
conjugate magnetic tail  fields
can be traced up to
$\approx$ 520 km above the photosphere. Above this height, the magnetic tails vanish because the magnetic field distribution becomes smoother, and moreover, the iPIL strongly shifts and is distorted with regard to its position and shape in the low photosphere.
Fig. \ref{fig:jmapflux2} shows an analog of Fig.\ref{fig:jmapflux1}, and the pertinent calculations were performed in the area enclosed by the green box
in Fig. \ref{fig:bz42}, but this time, located  at a height of 3 pixels above the photosphere.
As anticipated by the decrease with height of the magnetic field, the fluxes reported in this figure are lower than the corresponding photospheric
fluxes in Fig. \ref{fig:jmapflux1}. Moreover, and as 
also noted in Fig. \ref{fig:jmapflux1}, all displayed fluxes (i.e.  positive, absolute  negative, and
net unsigned) 
decrease at the same time, which is consistent
with magnetic flux cancellation. However,
we observe a delay of $\approx$ 7 time-steps (i.e.  608 s)  
in the onset of magnetic flux cancellation at $z=3$ pixels as compared to $z=0$ pixels. This delay is  anticipated because we used a flux-emergence simulation in which magnetic flux progressively reaches larger heights in the solar atmosphere. 

\subsection{Interpretation of magnetic flux cancellation}
In this section, we discuss the possible mechanism
that causes the flux cancellation in our simulation.
In Fig. \ref{fig:bzpilines} we display magnetic field
lines traced  from seeds within a sphere with a radius
of 2 pixels that is centred 2 pixels above the centre of the photospheric plane of the simulation box, which is  also the centre of the green box
in Fig.  \ref{fig:bz42}. Therefore, these  field lines essentially
permeate  the volume
in which magnetic flux cancellation takes place.
Fig. \ref{fig:bzpilines} shows two groups
of magnetic field lines: J-shaped lines rooted in the magnetic tail and spots,  and S-shaped field
lines that are sandwiched between the J-shaped lines and run along the iPIL. 
The J-shaped field lines owe their shape to the twisting and shearing photospheric motions discussed in the previous paragraphs,  which align them with the iPIL. They are  carried by the converging magnetic footpoint motions to the iPIL
(see, e.g. Fig.\ref{fig:montvh})
where they eventually reconnect and give rise to the S-shaped field lines. The transition
from J- to S-shaped field lines in PILs was frequently reported in both
observational and modelling studies following the work of \citet{1989ApJ...343..971V}, and 
we discuss this in more detail in Section 5. 
In tandem with the above,  the conjugate tail magnetic fields associated with the J-shaped field lines undergoing reconnection in the iPIL also 
give rise to the (photospheric) magnetic flux-cancellation signatures discussed in the previous
paragraphs. We therefore conclude that magnetic flux cancellation 
is associated with low-altitude magnetic reconnection between the J-shaped field lines.

To elaborate on the nature of the magnetic flux cancellation reported here, we first calculated the component of $V_z$ perpendicular to the magnetic field, $V_{z}^{\perp}$,
\begin{equation}
V_{z}^{\perp}=\frac{ |V_z \bold{k}  \times \bold{B}|}{|V_z||\bold{B}|} V_{z}  \bold{k},
\end{equation}
where $\bold{k}$ corresponds to the unit vector along the positive $z$ -axis. We opted to use  $V_{z}^{\perp}$ 
instead of $V_{z}$ as the former 
only corresponds to  the fraction of $V_{z}$ perpendicular
to the magnetic field. Therefore, $V_{z}^{\perp}$  can be viewed as a proxy of 
the vertical speed of the field lines and not of field-aligned plasma flows (see also
\citet{2012ApJ...754...15F} for a discussion of field line speeds with regard
to magnetic flux cancellation).  A positive (negative) $V_{z}^{\perp}$ implies that the corresponding
field line segment
rises (falls). 

We also calculated 
the  component of the magnetic tension vector
($ \sim \bold{B} \cdot \nabla \bold{B}$)
along the $z$ -axis, $\kappa_{z}$, normalised by the field magnitude,
\begin{equation}
\kappa_{z}=(B_x\frac{\partial B_z}{\partial x} +B_y\frac{\partial B_z}{\partial y} + B_z\frac{\partial B_z}{\partial z})/|\bold{B}|. 
\end{equation}
$\kappa_{z}$ with positive (negative) sign corresponds to field lines that are locally concave upward (downward).
Regions in photospheric PILs with positive $\kappa_{z}$ correspond
to the so-called bald patches \citep{1993A&A...276..564T}.

In panels (a) and (b) of Fig.\ref{fig:pil}, we show 
an overhead and a side view of
the S-shaped field lines in Fig. \ref{fig:bzpilines} coloured 
as a function of $V_{z}^{\perp}$ along their length.  We conclude that the S-shaped field lines generally correspond to positive 
$V_{z}^{\perp}$  with a magnitude of $\approx$ 0.2-0.5 $\mathrm{km \, s^{-1} }$. 
We assert
that the central parts of the S-shaped field lines, that is,  those corresponding
to  the green box in Fig. \ref{fig:bz42} in which magnetic
flux cancellation takes place, 
have positive $\kappa_{z}$.
Therefore,  from the analysis of
$V_{z}^{\perp}$ and $\kappa_{z}$, we conclude that 
the field lines slowly rise and are concave-upward in their middle in the
region
in which flux cancellation takes place. This is  equivalent to  emerging U-loops associated
with magnetic flux cancellation discussed in the Introduction.
In summary, and from the analysis of Figures  \ref{fig:bzpilines} and \ref{fig:pil}, we suggest that reconnection of the J-shaped field lines rooted
in the conjugate tails at either side of the iPIL gives rise to concave-upward slowly rising
magnetic field lines. The latter
explains the reduction in the photospheric magnetic flux that is associated with magnetic flux cancellation.

As discussed in the Introduction, 
magnetic reconnection 
associated with magnetic flux cancellation  can lead to
emerging (submerging) U- and $\Omega$- loops.
The emerging U-loops are  consistent with
the rising concave-upward field lines discussed in the previous paragraph. 
To provide a tentative explanation for the lack of submerging $\Omega$-loops, we followed \citet{1989ApJ...343..971V}. They
suggested that post-reconnection field lines in flux-cancellation sites can submerge when they are only short enough for the associated magnetic curvature to overcome 
the corresponding magnetic  buoyancy. This requirement is consistent with a maximum footpoint separation 
of $\approx$ 900 km, which is  a few times the photospheric scale-height, for submergence to occur. Assuming semicircular
field lines, we obtain an equivalent maximum field line length of
$\approx$ 1300 km. Finally, calculation of the lengths of the S-shaped field lines of Fig. \ref{fig:pil} gives values 
in the range $\approx$ 3300-4800 km,
which might explain why we did not obtain submerging field lines. 
This is confirmed in Fig. \ref{fig:jmapflux3}. This figure contains the time-series
of the positive, absolute negative, and net unsigned magnetic flux
for the green box in Fig. \ref{fig:bz42} in which we detected photospheric magnetic flux cancellation, this time  at a depth of 180 km below the photosphere. Fig. \ref{fig:jmapflux3}
clearly shows no increase in the relevant subphotospheric magnetic fluxes during the period when magnetic flux cancellation is observed in the photosphere. This increase would be expected if the cancellation were caused by the submergence of magnetic field lines.

\subsection{Magnetic flux cancellation and pre-eruptive MFRs}
We now  discuss the implications of our results in terms of pre-eruptive magnetic configurations. Following our discussion of
Fig. \ref{fig:bzpilines} in terms of the J- and S-shaped magnetic field lines, we note that the overall behaviour adheres to the so-called arcade-to-sigmoid transformation scenario originally advocated by \citet{1989ApJ...343..971V}, which
is a basic element of a number of observational
and theoretical and modelling works. According to this scenario, magnetic reconnection
converts J-shaped field (arcade) field lines into S-shaped (sigmoid) field lines, which  eventually leads to the formation of pre-eruptive MFRs.
The S-shaped field lines in Fig. \ref{fig:bzpilines}
do not correspond to a fully developed MFR, 
but they rather represent a seed thereof.
Additional reconnection in the iPIL
is required to generate truly twisted field
lines undergoing several crossings of the iPIL, and hence, eventually, of pre-eruptive MFRs as well
(see, e.g. the schematics and associated discussion by \citet{1989ApJ...343..971V} and \citet{2011A&A...526A...2G}).
This reconnection does not only lead to the formation of pre-eruptive MFRs, but also contributes to their subsequent growth (see also the pertinent MHD simulation results by \citep[e.g.][]{2010ApJ...708..314A, 2017ApJ...850...95S}).

To draw a more quantitative link between magnetic flux cancellation
and  pre-eruptive MFRs, we discuss  the pre-eruptive MFR in Fig. \ref{fig:bzjparlines}.  
This MFR corresponds to the yellow field lines that were traced from a quasi-circular $J_{par}$  concentration
in the $xz$-midplane outlined with the dashed line. The $J_{par}$  concentration used for the MFR field line tracing sits at the top of  a narrow vertical
$J_{par}$ concentration that is permeated with  
the purple magnetic field lines traced from the magnetic flux-cancellation region, as discussed for Fig. \ref{fig:bzpilines}.
We calculated the twist number,  $T_w$,  of these field lines,
\begin{equation}
T_w=\int_{L}\frac{(\nabla \times \vec{B}) \cdot \vec{B}}{4\pi B^2}dl,
\end{equation}
where the integration was performed along the length $L$ of each of the traced
field lines. $T_w$ essentially supplies the number of turns
that two infinitesimally adjacent field lines make around each other \citep{2006JPhA...39.8321B}.
We used the software described by \citet{2016ApJ...818..148L} in our calculations of $T_w$.
The average $T_w$ of the yellow field lines is  1.2 , and the corresponding
standard deviation is 0.5. Hence, this MFR is a weakly twisted MFR, as are most 
of the pre-eruptive MFRs reported by either observations or models \citep[e.g.][]{2020SSRv..216..131P}. 
We recall that
this MFR eventually erupted and gave rise to a blowout jet (see, e.g. the discussion of Fig. \ref{fig:jpar}).

We now calculated an estimate of the axial, that is,   associated with $B_{y}$,
magnetic flux, $\Phi_{y}$, of the pre-eruptive MFR. 
In calculating $\Phi_{y}$, we considered
pixels inside the $J_{par}$ concentration used to trace the MFR field lines.  The resulting $\Phi_{y}$ 
is  $4.2 \times {10}^{18} $ Mx, while the cancelled magnetic flux
in the interval encompassing the associated eruption and blowout jet, that is,  spanning time-steps
53 to 63 (see Fig. \ref{fig:ke}), is about $ {10}^{18} $ Mx.
This means  that magnetic flux cancellation might have roughly contributed up to $\approx$ the 25 $\%$
of the axial magnetic flux of the pre-eruptive MFR. Following the discussion of the previous
paragraphs, the magnetic flux associated with flux cancellation may have ended up 
in the pre-eruptive MFR as a consequence of multiple reconnections that lifted and coiled 
the original S-shaped field lines.  
For an in-depth analysis of the connection  between the magnetic fluxes associated
with magnetic flux cancellation and those sealed in pre-eruptive MFRs and its intricacies, we refer to \citet{2011A&A...526A...2G}.
It is obvious from the above
that the cancelled magnetic flux is neither  the sole nor the main contributor to the magnetic flux content of the pre-eruptive MFR. Reconnection and cancellation  above the photospheric
layers in which we observed magnetic flux cancellation might have been relevant, as shown, for example by \citet{2010ApJ...708..314A}, \citet{2017ApJ...850...95S}. As discussed previously,
beyond a certain height in the low photosphere, the magnetic tails vanish and  the magnetic field distribution becomes more homogeneous. This might in principle facilitate magnetic reconnection and magnetic flux cancellation along more extended segments in the respective iPILs. Calculating the contribution of reconnection and cancellation in multiple heights above the photosphere in the magnetic flux budget of pre-eruptive MFRs is a worthwhile future investigation.

\section{Conclusions and discussion}
We have presented a study 
of magnetic flux cancellation in a 3D MHD simulation
of the emergence of a twisted flux tube into a coronal hole that led 
to eruptions and jets.
Our main conclusions are summarised below. 
\begin{enumerate}
   \item We found photospheric magnetic flux cancellation
   between opposite-polarity and commensurate strength magnetic
 flux patches entrained in the magnetic tails of the emerging system.
\item The cancelling patches are associated with J-shaped (sheared) field lines, which converge and eventually reconnect along a short segment of the iPIL. The reconnection events generate slowly rising concave-upward  field lines that lead to the flux decrease associated with magnetic flux cancellation. 
   \item  The magnetic flux cancellation rate is $\approx$ $3.2 \times {10}^{18} $ $  \mathrm{Mx \, {hour}^{-1}}  $. Time periods when magnetic flux decreases  by 
15-20 \% were found during intervals encompassing individual eruptions and jets.   \item Magnetic flux cancellation is not temporally correlated with the standard jets of the simulation, but occurs when the system
     experiences eruptions and blowout jets.
   \item Magnetic flux cancellation can be traced to 520 km above the photosphere, with a delay of
   of around 10 minutes with respect to its occurrence in the photosphere. 
   \item Magnetic flux cancellation might have been relevant to the formation
   of MFR seeds, which
eventually grow and erupt and give rise to the blowout jets
   of the simulation. 
\end{enumerate}

Our simulations included
partial ionisation  of hydrogen in the low solar atmosphere. Whether this inclusion affects magnetic flux cancellation
in flux-emergence simulations requires further work.
We recall, however, that a handful of  MHD  simulations of emerging twisted flux tubes employing full ionisation of hydrogen indeed exhibited evidence
of magnetic flux cancellation as well, but no eruptive behaviour was reported \citep[][]{2011PASJ...63..417M,2012ApJ...754...15F}.

We first compared our results with the pertinent MHD flux-emergence
simulations of \citet{2011PASJ...63..417M} and \citet{2012ApJ...754...15F}.
As also discussed in the Introduction, both studies involved
the emergence of twisted flux tubes from the convection zone into the solar
atmosphere, and both gave rise to magnetic flux cancellation.  Similarly
to our results, these studies both involved slowly rising concave-upward field lines assuming the form of U-loops.
\citet{2011PASJ...63..417M} reported that the U-loops corresponded to
the bottom of the emerging flux tube, which can rise because their curvature is small.  \citet{2012ApJ...754...15F} reported that the U-loops
resulted from magnetic reconnection driven by convective and shear motions.  
Therefore, our results agree better with those by \citet{2012ApJ...754...15F}. 

We now compare our results with  observations.
We first compare the deduced magnetic cancellation
rate, that is,  $3.2 \times {10}^{18} $ $  \mathrm{Mx \, {hour}^{-1}}  $, with  pertinent observational results.
\citet{2018ApJ...853..189P} found  from HMI magnetograms an
average magnetic flux-cancellation rate $\approx$ $0.6 \times {10}^{18} $ $  \mathrm{Mx \, {hour}^{-1}}  $ for 13 ECH jets,  which
is significantly lower than our findings.
\citet{2016ApJ...823..110R} inferred magnetic flux-cancellation rates associated with Ellerman bombs inside
an active region in the range  $\approx 3.6 \times({10}^{17}-{10}^{18}) $ $  \mathrm{Mx \, {hour}^{-1}}  $.
\citet{2019A&A...622A.200K} found a magnetic flux-cancellation rate of $\approx 3.6 \times {10}^{18}$ $  \mathrm{Mx \, {hour}^{-1}}  $ in a young
active region.  \citet{2022ApJ...934...38L} found a magnetic flux-cancellation rate of $\approx 1.4 \times {10}^{18} $ $  \mathrm{Mx \, {hour}^{-1}}  $
for 38  QS magnetic flux-cancellation events. \citet{2022ApJ...924..137W} analysed 
1245 magnetic flux-cancellation events inside
an ECH  and found an average magnetic flux-cancellation rate 
of $\approx  {10}^{18} $ $  \mathrm{Mx \, {hour}^{-1}}  $. 
Interestingly, a common element in \citet{2016ApJ...823..110R}, \citet{2019A&A...622A.200K}, \citet{2022ApJ...934...38L} and \citet{2022ApJ...924..137W} that led to elevated magnetic flux-cancellation  rates which are  closer with our results  is that they all employed high-resolution magnetograms with a sub-arcesecond spatial resolution. We recall that the (horizontal) spatial resolution of our simulation of 180 km corresponds to scales that can be resovled with sub-arcsecond resolution observations, that is, they are below the spatial resolution of HMI. This implies that a simulation output at a spatial resolution commensurate to that of HMI could lead to lower magnetic flux-cancellation rates that are more consistent with HMI  observations.
The speed of the converging motions of the magnetic tails associated with magnetic flux cancellation of our simulation (i.e. $\approx$ 1 $\mathrm{km \, s}^{-1}$)
is consistent with  observations in magnetic flux cancellation sites  corresponding to speeds in the range 0.3-1.8 $\mathrm{km \, s}^{-1}$  
\citep[e.g.][]{2019A&A...622A.200K,2022ApJ...934...38L}.
\citet{2018ApJ...853..189P} found decreases of 21-73\% in the magnetic flux  during temporal intervals around 13 ECH jets; the lower value of their values is consistent with the largest 
decrease (that is  $\approx$ 20 \%) found in our simulation. 
Finally, our simulation results showing that blowout jets
occur only during periods of magnetic flux cancellation, are consistent with a number of observational
works discussed in the Introduction \citep{2017AGUFMSH43A2796M,2018A&A...619A..55M,2018ApJ...853..189P,2022ApJ...933...12M,2023ApJ...943...24P}. However, we have also discussed
observational works in the Introduction showing the contrary \citep{2019ApJ...873...93K,2021ApJ...909..133M}.
We note here that not
all of the jets of our simulation, and namely the standard jets, occurred when
magnetic flux cancellation was occurring. Summarizing, it seems as if our simulation
is consistent with several properties of magnetic flux cancellation and coronal jets and eruptions reported by observations.

An interesting result of our study is that magnetic flux cancellation
can be traced, with a  temporal delay of the order
of about 10 minutes, in multiple heights up to $\approx$ 520 km above the photosphere. The vertical extent of the magnetic cancellation region is associated
with the respective vertical extent of the magnetic tails.
Multi-height high-resolution observations of $B_{z}$, for example with the Daniel K. Inouye Solar Telescope (DKIST) 
\citep[e.g.][]{2022SoPh..297...22D} and with the Swedish Solar Telescope (SST)  \citet{2018ApJ...857...48G,2024ApJ...964..175G}  might
supply important tests of these predictions based on our modelling.

While our simulation is consistent with several basic aspects of 
magnetic flux-cancellation observations, it also gives rise to a major
difference. That is, the magnetic flux undergoing
cancellation (e.g. Fig. \ref{fig:jmapflux1}) is significantly lower by about a factor of 100 with respect
to the magnetic flux budget in the photospheric layer of
the entire simulation box (e.g. Fig. \ref{fig:ke}), which is essentially dominated
by the emergence of the twisted flux tube. 
This difference can readily be understood in terms 
of the small field of view of our calculations
of the magnetic flux cancellation, that is,  the green box in Fig. 
\ref{fig:bz42}. 
In retrospective, if we had used
a larger box than the green box in Fig. \ref{fig:bz42} to monitor the evolution of magnetic
flux, we would have missed the signature of magnetic flux cancellation, that is, the magnetic flux evolution
of Fig. \ref{fig:jmapflux1}. 

While opposite-polarity magnetic fields 
converge towards the iPIL along its entire length, with the exception of the fields inside the green box in Fig. 
\ref{fig:bz42} and for the most part of the considered temporal interval, they correspond to  largely disparate magnetic-field magnitudes, that is, to  spot-tail magnetic fields.
The interaction of these disparate magnitude fields may lead to the so-called unbalanced 
flux cancellation discussed by \citet{2007ApJ...666..576D}, which does not agree with the archetypal notion for flux cancellation, however, which involves opposite-polarity and similar-magnitude
magnetic fields. A potential application of unbalanced magnetic flux cancellation in our simulation could be found
during the standard jets in the external PIL of the system 
between the emerging and stronger negative flux (spot) and 
the ambient and weaker surrounding positive fields (e.g. see Fig. 3). Our calculation of the magnetic flux evolution in boxes bounding these fields only showed a decrease in the positive magnetic flux during the standard jets,  while the negative magnetic fluxes continued to increase. 
However, as amply demonstrated in the existing literature on  jets (see, e.g. the related literature in the Introduction)  magnetic reconnection above the photosphere is the key physical mechanism behind standard jets.
It therefore seems
rather hard to attribute the triggering of the (global) eruptions and jets of our simulation to magnetic flux cancellation.
This is corroborated by 
MHD simulations of larger-scale eruptions that 
suggest that the cancelled
magnetic flux is a significant fraction of the total available magnetic
flux, for example 6-11 \% in the studies by  \citet{2010A&A...514A..56A,2010ApJ...708..314A,2022ApJ...929L..23H}. The same constraint also follows from  observations \citep[e.g.][]{2018ApJ...853..189P}, showing that a sizeable fraction of the available magnetic flux  in sites in which coronal jets occur undergoes cancellation. Because magnetic flux
cancellation does not seem to be a plausible trigger mechanism of the jets and eruptions
of our simulation, other proposed mechanisms therefore need to be studied \citep[e.g.][]{2009ApJ...691...61P,2017Natur.544..452W}. This task is beyond the scope of our study, however. 

The important caveat of our simulation in terms of  the small fraction of the available magnetic flux involved in magnetic flux cancellation may be mitigated in future MHD modelling efforts
in at least two ways. First, by enhancing the magnitude of the ambient magnetic field in a number of  locations.
This  emulates unipolar enhanced magnetic field  patches in coronal holes, that is, an  enhanced network, which could cancel the minority polarity field of the emerging bipoles \citep[e.g.][]{2018ApJ...853..189P, 2022ApJ...933...12M}. 
Second, by considering the emergence and interaction of multiple twisted flux tubes
\citep[e.g.][]{2015ApJ...798L..10L, 2019ApJ...871...67C}. Both scenarios
might lead to the encounter and eventual annihilation of large patches of opposite-polarity magnetic fields, and might therefore increase
the fraction of the available magnetic flux undergoing magnetic flux cancellation. 
Finally, it will be worthwhile to investigate which properties of single emerging
twisted flux tubes such as geometry and twist distribution might be more prone
to lead to magnetic flux cancellation.

\begin{acknowledgements}

The
authors acknowledge support by the ERC Synergy Grant 'Whole Sun' (GAN: 810218). The authors extend sincere thanks to the referee for
useful suggestions and comments on the manuscript.

\end{acknowledgements}

\bibliographystyle{aa}
\bibliography{ref}

\end{document}